\documentclass[twoside]{dis07}
\usepackage[latin1]{inputenc}
\usepackage[dvips]{graphicx,epsfig,color}
\usepackage{wrapfig,rotating}
\usepackage{amssymb,amsmath,array}

\begin{document}
\title{Multi-Particle Decays of Light Mesons Measured by PHENIX at RHIC}

\author{Alexander Milov for the PHENIX Collaboration\footnote{for the
full list of authors see~\cite{ppg55}}
%
%
\vspace{.3cm}\\
%
Brookhaven National Laboratory - Dept. of Physics, \\
Upton NY 11973, USA
}

\maketitle

\begin{abstract}
The PHENIX experiment at RHIC measured $K^{0}_{S}$~$\eta$, and $\omega$-meson production at high $p_{T}$
in $p+p$, $d$+Au and Au+Au collisions at $\sqrt{s_{NN}}$ = 200~GeV. Measurements performed
in different hadronic decay channels give consistent results. This
paper presents measured meson-to-$\pi^{0}$ ratios and Nuclear
Modification factors in the most central $d$+Au and Au+Au
collisions. No suppression seen in $d$+Au interactions is in
contrast to a strong suppression of meson yields revealed in central
Au+Au collisions at the same energy.
\end{abstract}

\section{Analysis}
\begin{wrapfigure}{l}{0.63\columnwidth}
\centerline{\includegraphics[width=0.63\columnwidth]{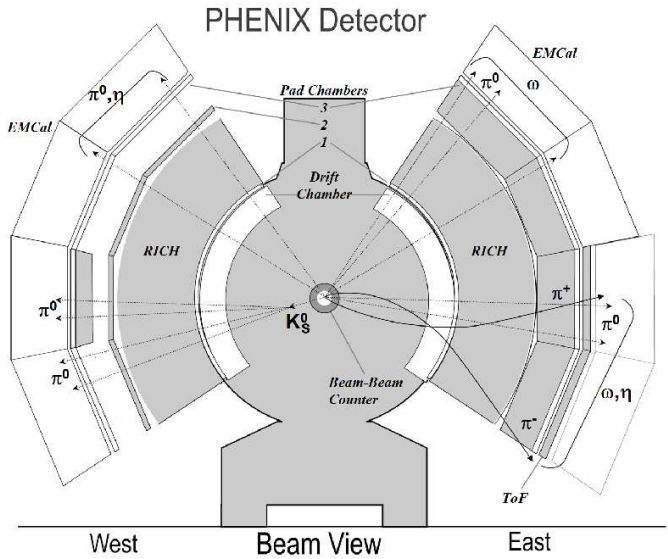}}
\caption{The PHENIX detector layout and the and the decays modes of
studied particles. PHENIX subsystems not used in the analysis are shown gray.}\label{fig:layout}
\end{wrapfigure}

The layout of the PHENIX detector~\cite{phenix} and the decay modes of the particles
presented in this analysis are shown in Fig.~\ref{fig:layout}. The
reconstruction begins with pairing photons (straight lines)
and selecting the $\pi^{0}$ candidates based on the invariant mass of the
pair. The $\pi^{0}$ candidates are then combined between themselves, other
photons or with the charged tracks and corresponding invariant mass distributions
are analyzed to extract the particle yields by simultaneous fitting of
the peaks and the background. The positions of the mass peaks were found
to be in agreement with the particle masses measured in vacuum
and the widths of the peaks, depending predominantly on the
detector resolution, change from 10~MeV/$c^{2}$ for
$\eta\rightarrow\pi^{0}\pi^{+}\pi^{-}$ to 20~MeV/$c^{2}$ for
$\omega\rightarrow\pi^{0}\pi^{+}\pi^{-}$ and
$K^{0}_{S}\rightarrow\pi^{0}\pi^{0}$ and to 30~MeV/$c^{2}$ for
$\omega\rightarrow\pi^{0}\gamma$. The values above vary within
less then 5 MeV/$c^{2}$ depending on the $p_{T}$ bin which agree with the widths
resulting from the detector resolution.

The analysis discussed in these paper is based on the event samples
accumulated during PHENIX physics Runs3,4, and 5 with integral
statistics, after quality assurance selection, corresponding to the
integrated luminosity of 1.5~nb$^{-1}$ ($p+p$) 129~$\mu$b$^{-1}$($d$+Au) and
2.5~pb$^{-1}$ (Au+Au) in these runs respectively. The background
conditions, depending on the mode of study for $p_{T}>$4-5~GeV/$c$, is
smaller than 1:5 to 1:20 in $p+p$ and $d$+Au and 1:70 in Au+Au.

\begin{figure}[h]
\centerline{\includegraphics[width=0.47\columnwidth]{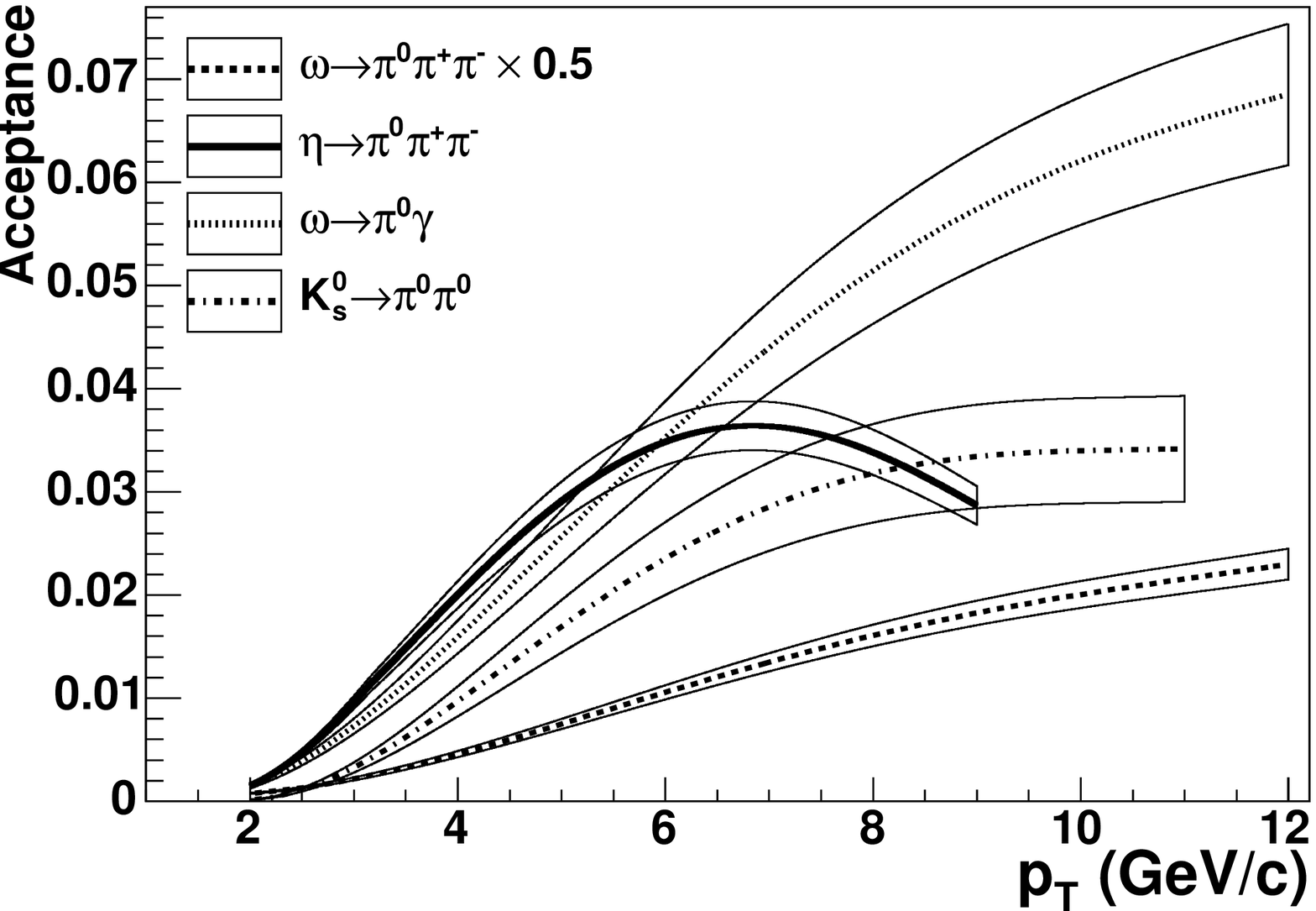}\includegraphics[width=0.47\columnwidth]{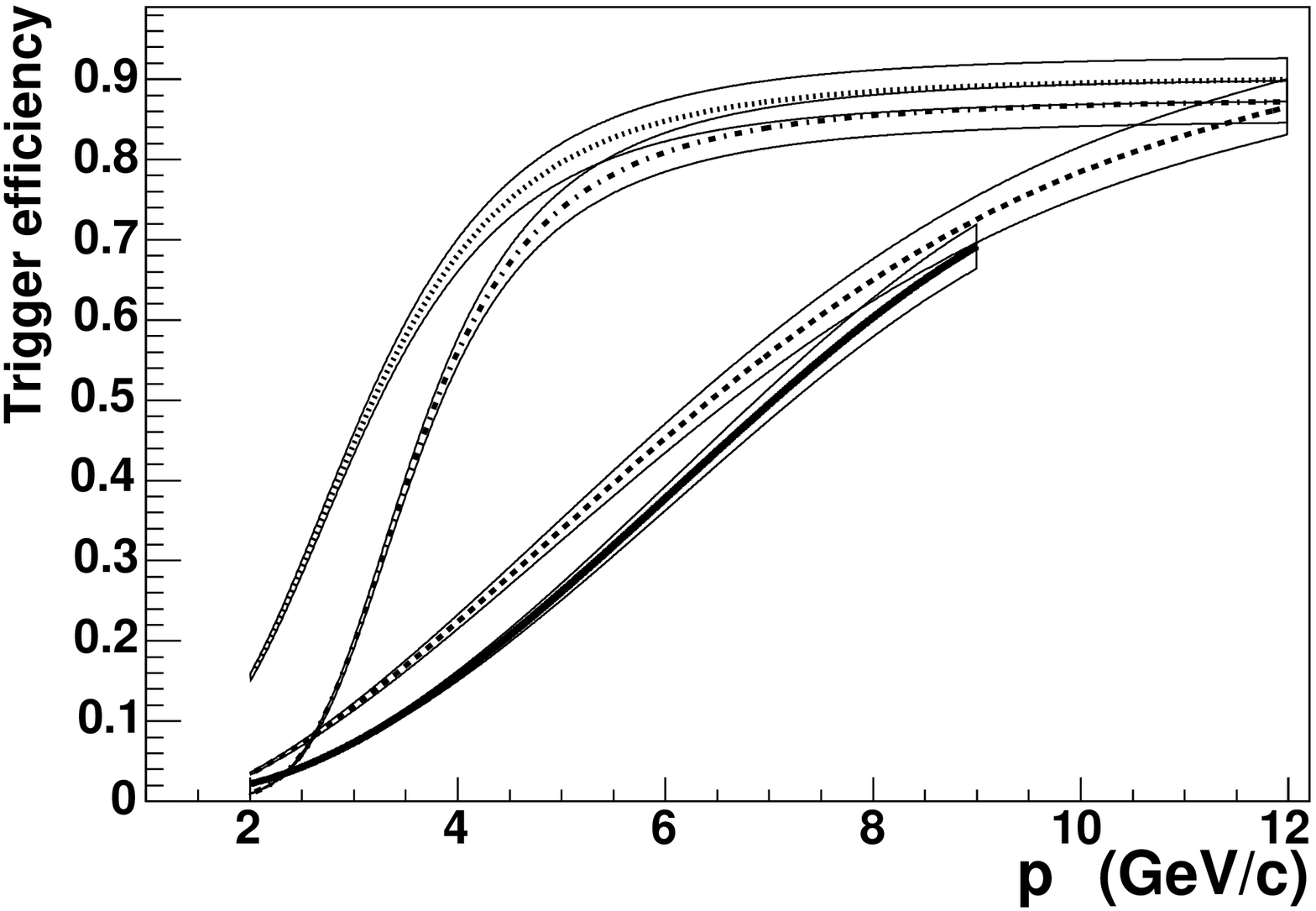}}
\caption{Shown on the left is the geometrical acceptance of the PHENIX detector for
various decay modes. The right panel shows the probability that a
photon coming from meson decay fires PHENIX high $p_{T}$ $\gamma$-trigger. A
unit corresponds to the standard ``Minimum Biased'' PHENIX trigger
registering 23$\pm$2.2~mb.}\label{fig:eff}
\end{figure}

The raw yields have to be corrected for the limited detector
acceptance, the $\gamma$-trigger efficiency, various
analysis cuts, the gamma conversions in the detector preceding the
calorimeter and the branching ratios of the specific decay mode. The
magnitude of the corrections is calculated based on the full detector
simulation and analysis of the data. The efficiencies measured
in the detector configuration used during the $p+p$ data taking are
shown in Fig.~\ref{fig:eff}.

The phase space density distribution of the true
three-body decay modes $\pi^{0}\pi^{+}\pi^{-}$, known from the
literature~\cite{phase} was taken into account in the simulation. More
details about this analysis can be found
in~\cite{ppg55,ppg64,victor_qm06}. For Au+Au events we applied an additional
correction for reconstruction losses due to the detector occupancy. 

The systematic errors of the measurement are listed in
Tab.~\ref{tab:err}. The main source of error is the extraction of
the raw yield made by fitting. The procedure is described in~\cite{ppg55}.
\begin{table}[h]
\centerline{\begin{tabular}{|l|cc|ccc|cc|}
\hline
Source & \multicolumn{2}{c|}{$\omega\rightarrow\pi^0\pi^+\pi^-$} 
       & \multicolumn{3}{c|}{$\omega\rightarrow\pi^0\gamma$} 
       & \multicolumn{2}{c|}{$K_{S}^{0}\rightarrow\pi^{0}\pi^{0}$} \\
& $p+p$  & $d+Au$ & $p+p$ & $d+Au$ & $Au+Au$ & $p+p$ & $d+Au$     \\ 
\hline
Acceptance             & $5 - 10$  & $9 - 12$   & $10 - 20$ & $8 - 12$ & $14 - 16$ & $10 - 25$ & $10 - 20$ \\
Trigger efficiency      & $3 - 10$  & $5 - 7$    & $2 - 7$   & 5           & -            & $2 - 10$  & 5            \\
Yield extraction       & $5 - 25$  & $10 - 15$  & $5 - 15$  & 10          & $15 - 35$ & $7 - 30$  & 9            \\
MB trigger             & 10           & 8                            & 10          & 8            & 4            & 10           & 8            \\
\hline
Total                  & $15 - 25$ & $18 - 22$  & $15 - 25$ & $17 - 20$& $20 - 45$ & $20 - 40$ & $15 - 25$ \\
\hline
\end{tabular}}
\caption{Systematic errors (in \%) for different decay modes and collision systems. Values with a range indicate
minimum and maximum error in the $p_{T}$ range of the measurement.}
\label{tab:err}
\end{table}

\section{Results}

\begin{wrapfigure}{l}{0.5\columnwidth}
\centerline{\includegraphics[width=0.5\columnwidth]{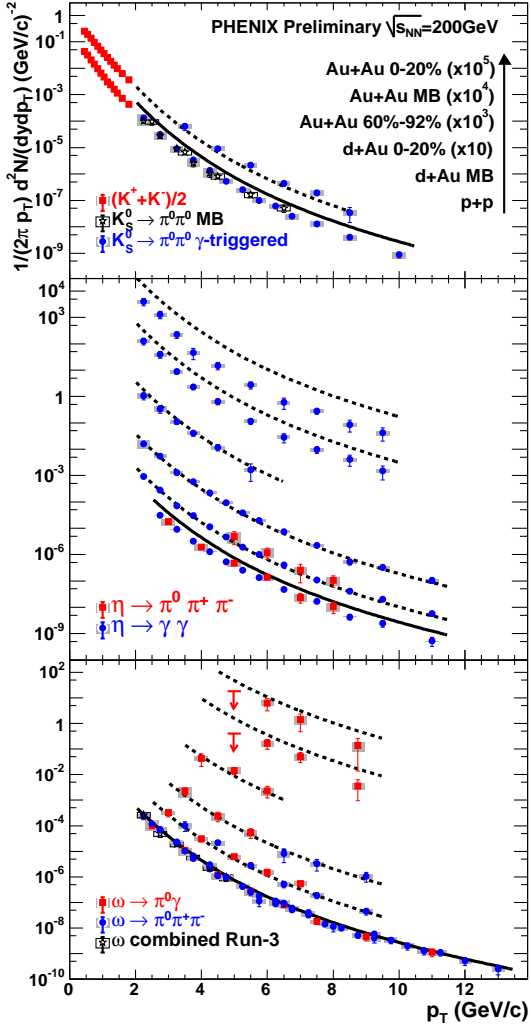}}
\caption{Invariant yields of K (top), $\eta$ (middle),
and $\omega$ (bottom) measured in $p+p$,
$p$+Au and Au+Au collisions at $\sqrt{s_{NN}}$=200~GeV. The solid line is
the parameterized $\pi^{0}$ spectra measured in $p+p$~\cite{pi0}. The dashed
lines are the $\pi^{0}$ spectra scaled by the
meson-to-$\pi^{0}$ ratio in $p+p$ and the number of the binary
collisions. K$^{+}$K$^{-}$ data is taken from~\cite{kpkm}}\label{fig:spectra}
\end{wrapfigure}

The results of the multi-particle decay measurements are presented in
Fig.~\ref{fig:spectra}
for K-meson in the top panel,
$\eta$-meson in the middle panel and $\omega$-meson in the bottom panel.

The $\omega$-meson is measured in two decay modes,
$\omega\rightarrow\pi^{0}\pi^{+}\pi^{-}$,
$\omega\rightarrow\pi^{0}\gamma$ and $\eta$-meson is measured in
$\eta\rightarrow\pi^{0}\pi^{+}\pi^{-}$, $\eta\rightarrow\gamma\gamma$.
The results of the measurements for the same meson agree. The
K$^{0}_{S}\rightarrow\pi^{0}\pi^{0}$ cannot be compared to
K$^{0}_{S}\rightarrow\pi^{+}\pi^{-}$ from PHENIX because of the
detector-induced background at the mass of the
K-meson. The results are in agreement with the STAR experiment
measurement in $\pi^{+}\pi^{-}$~\cite{star-ks}. We also see very good agreement
between the results obtained in PHENIX physics Run3 and Run5 and
between the results measured in triggered and Minimum Biased event
samples. In the latter case the correction shown in right panels of
Fig.~\ref{fig:eff} does not apply.

In the $p+p$ data the $p_{T}$ range of the measurement is limited by the detector
acceptance on the lower side of the range and by the available
statistics on the upper side. In Run5 the $\omega$-meson production in $p+p$
is measured out to 13~GeV/$c$ making it the second in $p_{T}$-reach
identified particle after the $\pi^{0}$. In the heavier collision
systems the combinatorial background effectively reduces the available statistics.

The lowest data points shown in each panel of Fig.~\ref{fig:spectra}
are the yields measured in $p+p$
collision. Plotted above them are the Minimum Biased and 0-20\%
central $d$+Au events data. The 60\%-92\% peripheral, Minimum Bias and
0-20\% central Au+Au collision spectra are plotted on top. For
$K^{0}_{S}$ such data are not available. All
measurements are done at $\sqrt{s},\sqrt{s_{NN}} = 200$~GeV. Central
$d$+Au and Au+Au data are scaled by different factors for the clarity.

The solid line shown in each panel is the parameterization of the
invariant yield of $\pi^{0}$ measured in $p+p$
collisions~\cite{pi0}. For the $\eta$- and the $K_{S}^{0}$-mesons this line
is above the data points and for the $\omega$-meson the points are
much closer. Using this parameterization the non-identical meson
ratios can be calculated. We find that in the $p+p$ collisions these
ratios are flat above the $p_{T}$=2.5~GeV/$c$. Fitted by a constant
the particle ratios are: $\omega/\pi^{0}$ = 0.81$\pm$0.02$\pm$0.07,
$\eta/\pi^{0}$= 0.48$\pm$0.02$\pm$0.02 and $K^{0}/\pi^{0}$ = 0.45$\pm$0.01$\pm$0.05.

Each set of points shown in Fig.~\ref{fig:spectra} has a corresponding
dashed line. These lines are constructed in the following way: the
$\pi^{0}$ spectra measured in $p+p$ (solid line) is scaled with the corresponding
meson-to-$\pi^{0}$ ratio given above. Since all ratios are found to be
flat in the region of the measurement, the scaled $\pi^{0}$ reference
corresponds to the invariant yield of the meson in the $p+p$
collisions. Those yields are further scaled by the number of binary
collisions for each centrality bin in $d$+Au and Au+Au presented in
the figure.

As one can see for all analyzed mesons the $d$+Au data in
Minimum Bias and 0-20\% most central events are very close to the
dashed lines. The ratio of the two is the Nuclear Modification Factor and for
the analyzed mesons it was found to be flat within the errors of the
measurement. For the Minimum Bias event sample the $R_{dA}$'s are
above unit but agree with 1 within the errors of the measurement.

In the peripheral Au+Au collisions the nuclear modification factor is
not very different from 1 as the dashed line lays close to the
points. This is not so in the Minimum Biased and 0-20\% central Au+Au
events. For $\omega$-meson we find the $R_{AA}$ to be 0.4$\pm$0.15. The
$\eta$-meson production in central Au+Au events is suppressed by a
factor of 5 compared to scaled $p+p$ reference.

\section{Acknowledgments}
Work of the speaker is supported by the
Goldhaber Fellowship at BNL with funds
provided by Brookhaven Science Associates.


\end{document}